\documentclass[oldversion]{aa}
\pdfoutput=1

\usepackage{amsmath}
\usepackage{txfonts}
\usepackage[american]{babel}
\usepackage{graphicx}
\usepackage{xspace}

\usepackage{natbib}
\newcommand{\Alfven}{Alfv\'{e}n\xspace}
\newcommand{\Alfvenic}{Alfv\'{e}nic\xspace}
\newcommand{\bnd}{\ensuremath{\mathrm{b}}}
\newcommand{\Nabla}{\vec{\nabla}}

\newcommand{\pmag}{\ensuremath{p_\text{mag}}}
\newcommand{\vA}{\ensuremath{v_\mathrm{A}}}
\newcommand{\vAr}{\ensuremath{v_{\mathrm{A}r}}}
\newcommand{\cs}{\ensuremath{c_\mathrm{s}}}
\newcommand{\cso}{\ensuremath{c_{\mathrm{s}0}}}
\newcommand{\evarphi}{\ensuremath{\hat{\vec{e}}_\varphi}}
\newcommand{\de}{\ensuremath{\mathrm{d}}}
\newcommand{\const}{\ensuremath{\text{const.}}}
\newcommand{\degree}{\ensuremath{^\circ}}
\newcommand{\mfrate}{\ensuremath{\mathcal{M}}}
\newcommand{\efrate}{\ensuremath{\mathcal{E}}}
\newcommand{\mean}[1]{\ensuremath{\langle#1\rangle}}
\newcommand{\poloidal}{\ensuremath{\mathrm{p}}}
\newcommand{\Ma}{\ensuremath{M_\mathrm{A}}}

\begin{document}

\title{Decay of the toroidal field in magnetically driven jets}
\author{R. Moll}
\offprints{\protect\raggedright R. Moll, \email{rmo@mpa-garching.mpg.de}}
\institute{Max-Planck-Institut f\"{u}r Astrophysik, Karl-Schwarzschild-Str. 1, 85748 Garching, Germany}
\date{}
\abstract{ A 3D simulation of a non-relativistic, magnetically driven jet
propagating in a stratified atmosphere is presented, covering about three
decades in distance and two decades in sideways expansion.  The simulation
captures the jet acceleration through the critical surfaces and the development
of (kink-)instabilities driven by the free energy in the toroidal magnetic
field component.  The instabilities destroy the ordered helical structure of
the magnetic field, dissipating the toroidal field energy on a length scale of
about 2--15 times the \Alfven distance.  We compare the results with a 2.5D
(axisymmetric) simulation, which does not become unstable.  The acceleration of
the flow is found to be quite similar in both cases, but the mechanisms of
acceleration differ. In the 2.5D case approximately 20\% of the Poynting flux
remains in the flow, in the 3D case this fraction is largely dissipated
internally. Half of the dissipated energy is available for light emission; the
resulting radiation would produce structures resembling those seen in
protostellar jets.  }
\maketitle

\section{Introduction and rationale of the calculations}

A magnetized outflow produced by a rotating magnetic object has become the
default interpretation for objects ranging from protostellar jets to gamma-ray
bursts. The outflow in this model contains a tightly wound helical magnetic
field. Such a nearly toroidal field represents a source of free energy that
makes the flow inherently prone to non-axisymmetric magnetic instabilities.  In
this study we investigate the longer-term development of such instabilities and
their consequences for jet phenomenology.

Instabilities are not necessarily fatal for the jet. Kink instabilities of
helical magnetic field configurations typically saturate at a finite amplitude.
Such instabilities at moderate amplitudes have been invoked to explain
phenomena like the wiggly appearance of Herbig-Haro objects \citep{1993Todo} or
the orientation of VLBI jets in AGN \citep{1985Koenigl}. 

The development of kink instability in a jet is not the same as in a laboratory
configuration.  Due to the sideways expansion of the flow, the ratio of
poloidal (stabilizing) to toroidal (destabilizing) field components decreases
with distance along the flow. Conditions favorable for instability are thus
continually recreated in such a flow.  In \citet{2008Moll}, hereafter Paper~1,
we presented 3D MHD simulations showing the onset of instabilities in an
expanding jet created by twisting a purely radial magnetic field.  The degree
of instability was found to depend on the kind of rotation that generates the
twist.  The highest degree of instability was attained with a constant angular
velocity (rigid rotation); the jet produced in this case was subject to helical
deformations with large amplitudes, causing sideways displacements of several
degrees.  However, the magnetic structure of the jet was not disrupted within
the computational volume, and the instability did not lead to a significant
decrease of the Poynting flux.  These results indicate the need to follow the
instabilities to larger distances from the source. We present here the results
of simulations extending to a distance of 1000 times the diameter of the jet
source.

In Paper~1 we showed how the degree of instability depends on the way in which
the jet is collimated by its environment. If collimation conditions are such
that the opening angle of the jet increases with distance, the \Alfven travel
time across the jet increases more rapidly than the expansion time scale. As a
result, instabilities soon ``freeze out'', and decay of the toroidal field
becomes ineffective. In better collimating jets, such that the opening angle
narrows with distance, instability is  always effective. The calculations
presented here are for such a case. It is probably the most relevant for both
AGN and protostellar jets. Observations of the jets in M87 \citep{1999Junor}
and HH30 \citep{1990Mundt}, for example, show a rapidly decreasing opening
angle in the inner regions of the jet.

The dissipation of the magnetic energy  in the toroidal field heats the plasma.
Radiation produced by this plasma would be an alternative or complement to the
standard mechanism invoked, which relies on dissipation in internal shocks.
Observations such as those of M87 and HH30 indicate that dissipation starts
fairly close to the central object, compared with most observable length
scales, but also that  the innermost regions, perhaps comparable to the \Alfven
distance $r_\mathrm{A}$, are quiet. Inferences from VLBI observations  indicate
that AGN jets are not magnetically dominated on most observable scales
\citep[cf.][and references therein]{2005Sikora}; decay of the magnetic field
relatively close to the source of the jet would fit this observation \citep[see
also discussion in][]{2006Giannios}.

If dissipation takes place close enough to the source, it is possible that it
can be studied realistically by 3D simulations with current computational
resources. In Paper~1, we already found indications that decay of the toroidal
field may become important at distances as close as 10--30 times
$r_\mathrm{A}$.

The release of magnetic energy by the instability may  also be  important  for
accelerating the flow \citep{2002Drenkhahn}.  Dissipation of toroidal field
causes the magnetic pressure gradient along the jet to steepen; this adds an
accelerating force that is absent when the toroidal field is conserved
\citep{2006Giannios,2008Spruit}.

To facilitate extraction of physics from the numerical results,  the 3D results
in the following are compared systematically with a 2.5D (axisymmetric)
simulation corresponding to the same initial and boundary conditions.

\section{Methods}
\label{sec:methods}

\subsection{Numerical MHD solver, grid and coordinates}
\label{sec:nummethods}

The following is a brief summary; for details see Paper~1 where the same
numerical approach but different initial conditions were used.  Details on the
MHD code can also be found in \citet{2008Obergaulinger}.

We numerically solve the ideal adiabatic MHD equations in a static external
gravitational potential $\Phi \propto r^{-1}$ on a spherical grid
$(r,\theta,\phi)$.  In the 2.5D, axisymmetric simulation, the jet propagates
along the coordinate axis $\theta=0$.  In the 3D simulation, the jet's  axis is
taken in  the direction $\theta=\phi=\pi/2$.  The jet thus propagates in
equatorial direction of the coordinate system. This avoids the coordinate
singularity along the polar axis ($\theta=0$), which is numerically
problematic for trans-axial flows such as are caused by instabilities.  The
computational volume covers a range $\Delta\theta=\Delta\phi$ that comprises
about twice the expected opening angle of the jet.  The spacing of the
computational grid is uniform in the angular directions and logarithmic in the
radial direction. In this way, the varying numerical resolution approximately
matches the increase of the natural length scales in the expanding jet.

For presentation and discussion, we transform the results to a different
coordinate system. This is again a spherical coordinate system, but with the
polar axis aligned with the jet.  The polar and azimuthal angles  in this
system are denoted by $\vartheta$ and $\varphi$, respectively:
\begin{align}
    \sin^2\vartheta &= 1-\sin^2\theta \sin^2\phi, \\
    \tan\varphi &= \tan\theta \cos\phi
\end{align}
in the 3D simulation and $\vartheta \equiv \theta$, $\varphi \equiv \phi$ in
the 2.5D simulation.  $R \coloneqq r\sin\vartheta$ is used to denote the
distance to the axis (cylindrical radius).  

Dissipation of magnetic energy in the flow causes heating. In nature this would
lead to losses by radiation; in the computations it can cause numerical
problems in regions where the magnetic energy dominates. Instead of a more
realistic model for such losses,  a temperature-control term is added in the
energy equation, such that the temperature relaxes to that of the initial state
on an appropriate time scale. In the simulations presented here this time scale
is chosen such that the temperature stays within about a factor 100 around the
initial value.  

\subsection{Initial and boundary conditions}
\label{sec:iandbcs}

The initial state consists of a current-free magnetic field embedded in a
stratified atmosphere in the gravitational field of a point mass. The field
configuration of this initial state is of a ``collimating'' type, the distance
between neighboring field lines increases less rapidly than the distance from
the source $r$.  In Paper~1, we showed that instabilities develop more strongly
in such collimating configurations than in a purely radial initial field.

\subsubsection{Initial field configuration}

The initial field is  axisymmetric around the jet axis and hence can be written
as
\begin{equation}
    \vec{B} = \frac{1}{R} \Nabla \psi \times \evarphi = \Nabla \times \left( \frac{1}{R} \psi \evarphi \right)
            = \Nabla \times \vec{A},
\label{}
\end{equation}
where $\psi$ is the stream function of the field. We construct the initial
condition as a linear combination of the stream function of a monopole field,
given by
\begin{equation}
    \psi_\text{mono} \propto  1 - \cos\vartheta ,
\label{}
\end{equation}
and a field with the stream function
\begin{equation}
    \psi_\text{para} \propto \sqrt{ \left( \frac{R}{R_0} \right)^2
                       + \left( 1 + \frac{z}{R_0} \right)^2 }
                       - \left( 1 + \frac{z}{R_0} \right),
\label{}
\end{equation}
($z>0$)  of which the field lines have a parabolic shape \citep{1994Cao}.
Here, $z=r\cos\vartheta$ denotes the height along the jet's central axis (which
is actually the $y$-axis in the 3D simulation and the $z$-axis in the 2.5D
simulation).  The weighting of the two components  is parametrized with the
quantity
\begin{equation}
    \zeta \coloneqq \frac{B_\text{mono}}{B_\text{para}} \Biggr|_{r=r_\bnd,\vartheta=0},
\label{}
\end{equation}
the relative strength of the constituent fields at the lower boundary
$r=r_\bnd$.  $\zeta \rightarrow\infty$ corresponds to a pure monopole field
and $\zeta = 0$ corresponds to a pure parabolic field. The radius
$r_\text{equ}$ at which $B_\text{mono}=B_\text{para}$ on the central axis
follows from
\begin{equation}
    \zeta = \frac{r_\text{equ}^2(R_0+r_n)}{r_n^2(R_0+r_\text{equ})} .
\label{}
\end{equation}
The lower the value of $\zeta$ or $r_\text{equ}$, the more collimating is the
shape of the initial field. 

\subsubsection{Stratification}

We impose the static gravitational field of a point mass, located at the origin
of the coordinate system. The stratification is initially in hydrostatic
equilibrium in this potential.  The gas pressure is chosen such that the
plasma-$\beta \coloneqq p / \pmag$ in the initial state is approximately
constant at small radii ($\mathord{\ll}r_\text{equ}$), where the monopole field
dominates: $p \propto r^{-4}$, $\rho\propto r^{-3}$, $T \propto r^{-1}$, $\cs
\propto r^{-1/2}$ and $\vA \propto r^{-1/2}$ in the initial state.  At large
radii, the value of the initial $\beta$ is reduced as the magnetic field
strength decreases less rapidly than $r^{-2}$.

\subsubsection{Boundaries}

Boundary conditions are maintained through the use of ``ghost cells'' outside
the computational domain.  At the sides ($\theta$ in the 2.5D simulation,
$\theta$ and $\phi$ in the 3D simulation) and top (upper $r$), we use open
boundary conditions that allow for an almost force-free outflow or inflow of
material, including magnetic fields.

The  bottom boundary is located at a finite height $r_\bnd$ above the origin of
the gravitational potential. This distance is 1/200 of the size of the
computational volume. The jet is generated there by a  ``rotating
disk''\footnote{Strictly speaking, because of the spherical grid, what we call
``disk'' here is really a spherical cap with small curvature.} of radius
$R_\bnd$ around  the axis. It is implemented by maintaining a velocity field
$\vec{v}=v_\varphi \evarphi$ in the ghost zones:
$v_\varphi=v_{\varphi,\bnd}^\text{max} R/R_\bnd$ for $R \le R_\bnd$ and $0$
elsewhere. The disk thus rotates rigidly like in the cases R2 and R3 in
Paper~1.

All quantities at the bottom boundary except for $\vec{B}$ are fixed at their
initial values in the ghost cells. $\vec{B}$ is extrapolated from the interior
of the domain.

\subsection{Parameters and units}

\begin{table}
\caption{Normalization units}
\centering
\begin{tabular}{ccc}
\hline\hline
Quantity & Symbol(s) & Unit \\
\hline
    length          & $x$,$y$,$z$,$r$,$R$   & $l_0$ \\
    gas pressure    & $p$                   & $p_0$ \\
    density         & $\rho$                & $\rho_0$ \\
    mass flow rate  & $\mfrate$             & $\rho_0 l_0^3 / t_0$ \\
    velocity        & $v$                   & $\cso = \sqrt{ \gamma p_0 / \rho_0 }$ \\
    time            & $t$                   & $t_0 = l_0 / \cso$ \\
    energy          & $E$                   & $p_0 l_0^3$ \\
    energy flow rate& $\efrate$             & $p_0 l_0^3 / t_0$ \\
    force density   & $F$                   & $p_0 / l_0$ \\
    power           & $P$                   & $P_0 = p_0 l_0^3 / t_0$ \\
    magnetic flux density       & $B$       & $B_0 = \sqrt{8 \pi p_0} $ \\
    current density  & $j$                   & $j_0 = B_0 c / (4\pi l_0)$ \\
\hline
\end{tabular}
\label{tab:units}
\end{table}

The above suggests a specification of the problem in terms of 7 parameters: the
field configuration parameter $\zeta$, a scale for the field strength, scales
for the pressure and density, the bottom boundary location  $r_\bnd$, the
radius $R_\bnd$ of the rotating disk, and its rotation rate.  Because of the
symmetries of the problem, it is actually determined by only 4 parameters; the
remaining 3 dependences are equivalent to scaling factors.  As the 4
independent parameters we choose the following dimensionless quantities: the
plasma-$\beta$ of the initial state, the angle
$\vartheta_\bnd=\arctan(R_\bnd/r_\bnd)$ which controls the opening
angle of the flow, the  field configuration parameter $\zeta$, and finally the
\Alfvenic Mach number $\Ma$  of the rotating velocity field at the edge
($R=R_\bnd$) of the launching disk, which controls the power of the jet.
They have the values $\beta=1/9$, $\zeta=30$, $\vartheta_\bnd=5.7\degree$,
$\Ma=0.1$.  [The simulations reported in Paper~1 corresponded to  $\beta=1/9$,
$\zeta\rightarrow\infty$, $\vartheta_\bnd=5.7\degree$, $\Ma=0.1$.]

The units used for reporting the results below are the length scale $l_0 \equiv
2 R_\bnd$, and the pressure and density on the axis at the bottom boundary,
$p_0 \equiv p_\bnd$, $\rho_0 \equiv \rho_\bnd$. Together with the 4 model
parameters $\beta$, $\zeta$, $\vartheta_\bnd$ and $\Ma$, the units for
other quantities follow from these as shown in Table~\ref{tab:units}.  

The simulations cover a distance of 2000 times the initial jet radius
$R_\bnd$ in the spherical range $5<r<1005$. The resolution used in the 3D
simulation is $768 \times 128 \times 128$; the corresponding domain size is
$1000 \times 33.8\degree \times 33.8\degree$. The resolution used in the 2.5D
simulation is $768 \times 96$;  the  corresponding domain size $1000 \times
16.8\degree$. In both simulations, the radial width $\Delta r$ of the grid
cells increases from $0.03$ at the lower boundary to $6.92$ at the upper
boundary in logarithmic steps. 

\section{Results}
\label{sec:results}

\begin{figure*}[t]
\begin{center}
\includegraphics[width=.62\linewidth]{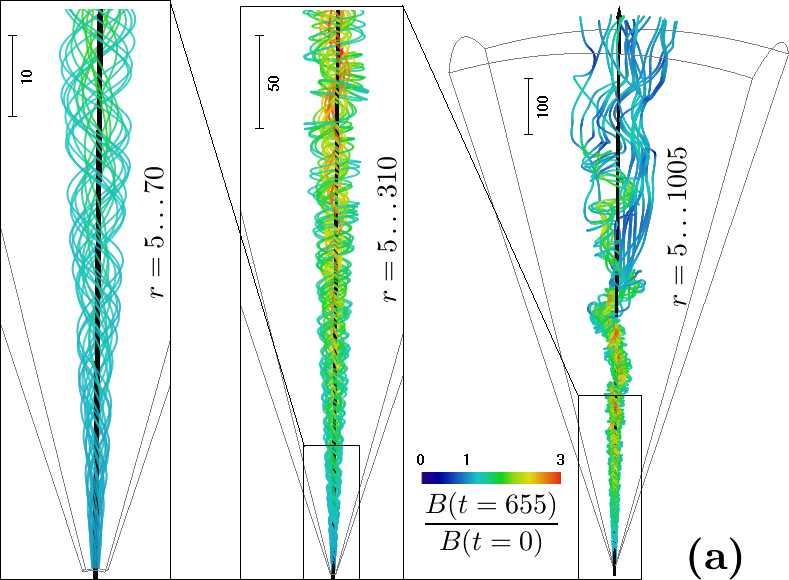}
\includegraphics[width=.368\linewidth]{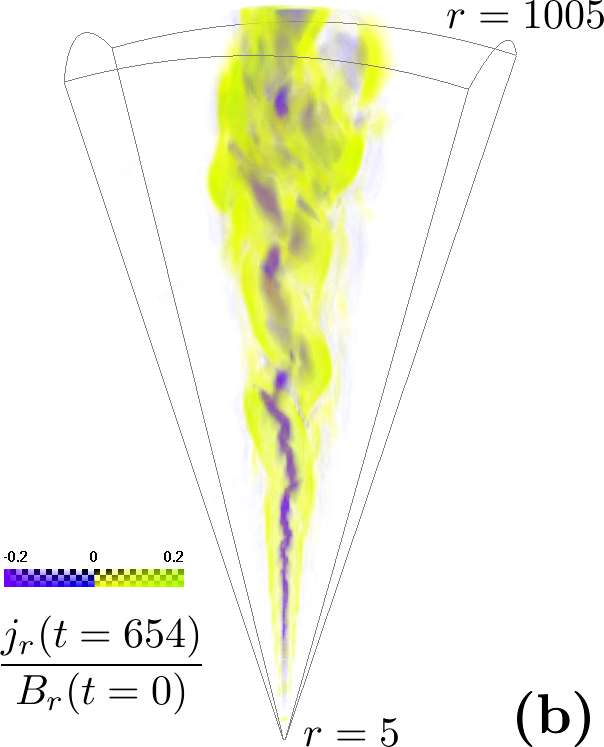}
\end{center}
\caption{\textbf{a)} Selected magnetic field lines in the 3D simulation on
successively increasing length scales.  The color coding gives the magnetic
field strength relative to its initial value. The field lines shown are the
ones that are anchored in the rotating disk at the lower boundary.  The jet
starts out with a helical magnetic field ({\it left image}) whose toroidal
component becomes increasingly stronger ({\it middle image}) until
instabilities disrupt the ordered structure, the toroidal field decays and the
field becomes predominantly poloidal ({\it right image}).  \textbf{b)} Radial
component of the current density ($\Nabla \times \vec{B}$) in the 3D
simulation.  The backward current, shown in blue here, is concentrated on the
central axis until being first displaced and then disrupted by instabilities.}
\label{fig:bfield}
\end{figure*}

\begin{figure}[t]
\begin{center}
\includegraphics[width=\linewidth]{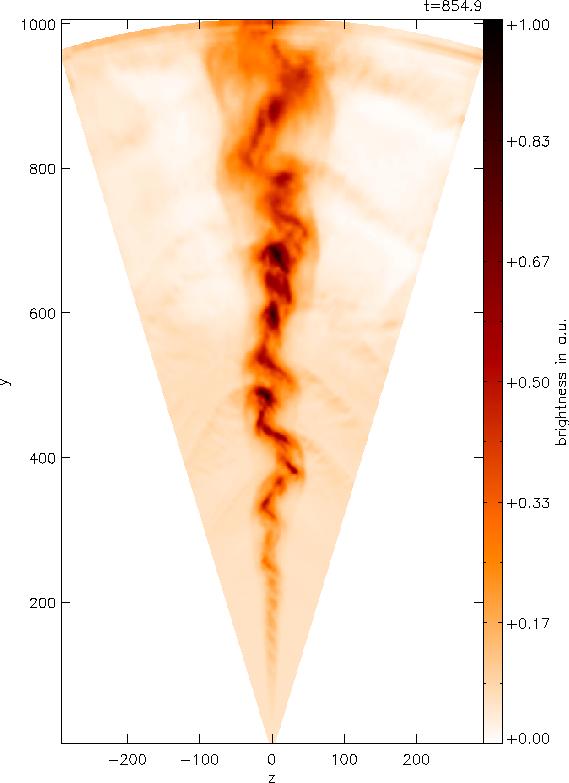}
\end{center}
\caption{Volume rendered image of the 3D jet that shows what it might look like
in observations.  For the volume rendering, a simple model was used in which
emissivity and opacity depend on temperature and magnetic field strength. The
jet exhibits a wiggly structure with bright knots, produced by the
instabilities and the dissipation of magnetic energy.}
\label{fig:tvolren}
\end{figure}

The 3D simulation was done using MPI parallelization on 128 CPUs, the 2.5D
simulation was done with OpenMP parallelization on 32 CPUs.

The jet crossed the upper boundary of the computational volume at the physical
time $t \approx 523$ on the 13th wall clock day of the 3D simulation. We
stopped it after 26 days, at which time $t = 1055$ had been reached.  The 2.5D
simulation ran for 24 hours, reaching $t=1732$.  

The 3D jet is subject to non-axisymmetric instabilities, evidently of the kink
($m=1$) kind. They have a disruptive effect on the magnetic field structure and
cause the toroidal field to decay, see Fig.~\ref{fig:bfield}.  If such a jet
was observed, it would probably look similar to Fig.~\ref{fig:tvolren}, with
bright knots and wiggles being prominent features. The knots move at a
substantial fraction of the flow speed, sometimes merging or fading before
leaving the computational domain.  The 2.5D jet does not exhibit any form of
instability.

\subsection{Acceleration, collimation and mass flow}

\begin{figure}[t]
\begin{center}
\includegraphics[width=\linewidth]{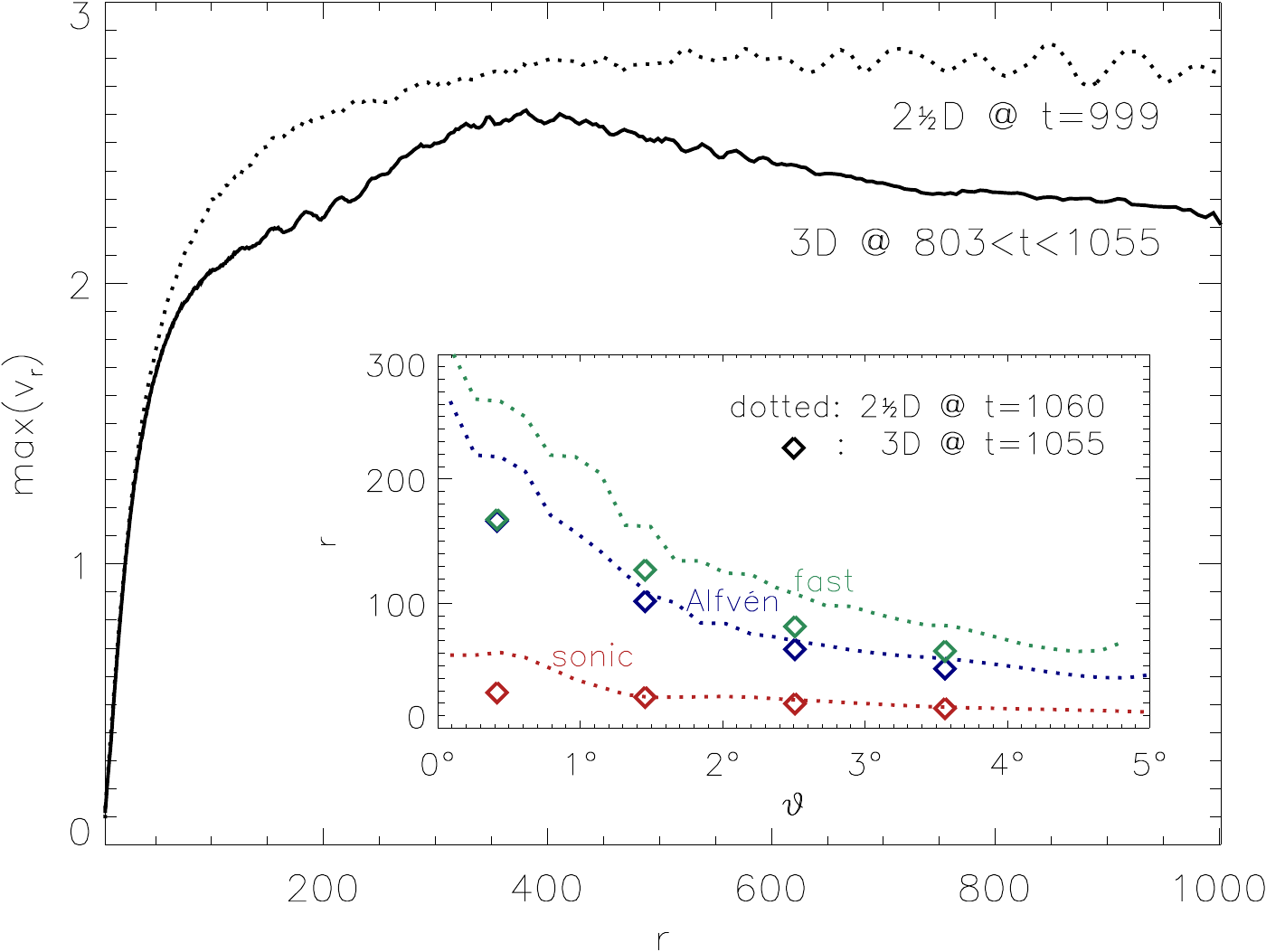}
\end{center}
\caption{Maximum velocity in jet direction as a function of distance. The inset
shows the location of the critical surfaces. The flow passes first the sonic,
then the \Alfven and finally the fast magnetosonic surface.}
\label{fig:velocity}
\end{figure}
\begin{figure}[t]
\begin{center}
\includegraphics[width=\linewidth]{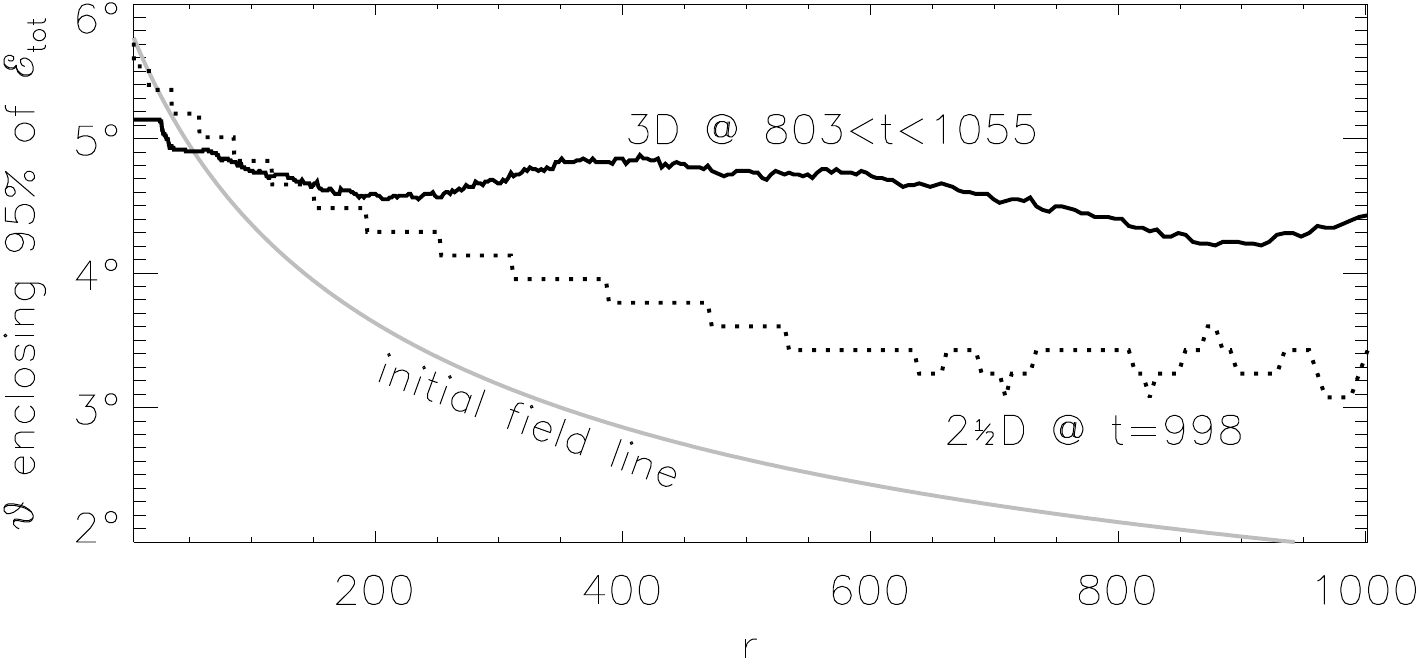}
\end{center}
\caption{Jet boundary, defined by the angle that encloses 95\% of the total
energy flow (see Eq.~\ref{eq:eflux}), as a function of distance.  The  unstable
3D jet is less collimated than the 2.5D jet.}
\label{fig:border}
\end{figure}

Most of the acceleration takes place below $r=100$, where the jet reaches
$\mathord{\gtrsim}80\%$ of its terminal speed, see Fig.~\ref{fig:velocity}. The
location of the sonic ($v_r=\cs \approx \cs\vA/(\cs^2+\vA^2)^{1/2}$, the slow
magnetosonic cusp speed), \Alfven ($v_r=\vAr$) and fast magnetosonic
($v_r^2=c_s^2+v_A^2$) radii depends on the direction $\vartheta$, see inset in
Fig.~\ref{fig:velocity}.  This is because the acceleration is more effective
near the boundary of the jet, which is a consequence of the rigid rotation
profile used in its generation, and because the poloidal magnetic field is
being redistributed such that the local \Alfven velocity becomes relatively
high near the axis.  As the toroidal field energy, which determines the jet
acceleration and the development of instabilities, is concentrated towards the
jet boundary (as opposed to the axis), the \Alfven radius there is probably the
most important for the subsequent considerations.

The central velocity tends to be higher in the 3D simulation, presumably due to
a more effective transfer of momentum from the boundary to the center.  This
may, together with the entrainment of ambient material discussed below, also
explain why the peak velocities are somewhat lower in the 3D simulation.

The instabilities have a noticeable effect on the collimation behavior, see
Fig.~\ref{fig:border}.  In the 2.5D simulation, the opening angle of the jet
decreases by about $2\degree$ in the first half of the computational volume,
which is about $1\degree$ less than what is marked by the shape of the initial
magnetic field.  In the second half, well beyond the \Alfven surface, the
opening angle settles to a constant value.  The jet in the 3D simulation is
less collimated\footnote{Note, however, that with our definition of the jet
boundary, a rigid displacement away from the center also increases the opening
angle.}, the location of the boundary fluctuates with time in the unstable
region.  Averaged over time, we find it to be nearly conical. 

The mass flow rate $\mfrate(t,r) \coloneqq \int_{r=\const} \rho v_r \, \de A$
is somewhat higher in the 3D simulation and subject to strong fluctuations
above $r\approx 200$: the mean value over all radii at $t\approx1060$, measured
in the unit for $\mfrate$ listed in Table~\ref{tab:units}, is $0.082\pm34\%$ in
the 2.5D and $0.082\pm34\%$ in the 3D simulation.  The average over several
time steps shows a slight increase of $\mfrate$ with $r$ in the 3D case,
whereas no trend can be deduced in the 2.5D case.  This may be an indicator for
an enhanced entrainment of ambient material caused by the instabilities.  It
may also in part explain the lower peak velocities in the 3D simulation.

\subsection{Energy}
\label{sec:energy}

\begin{figure}[t]
\begin{center}
\includegraphics[width=\linewidth]{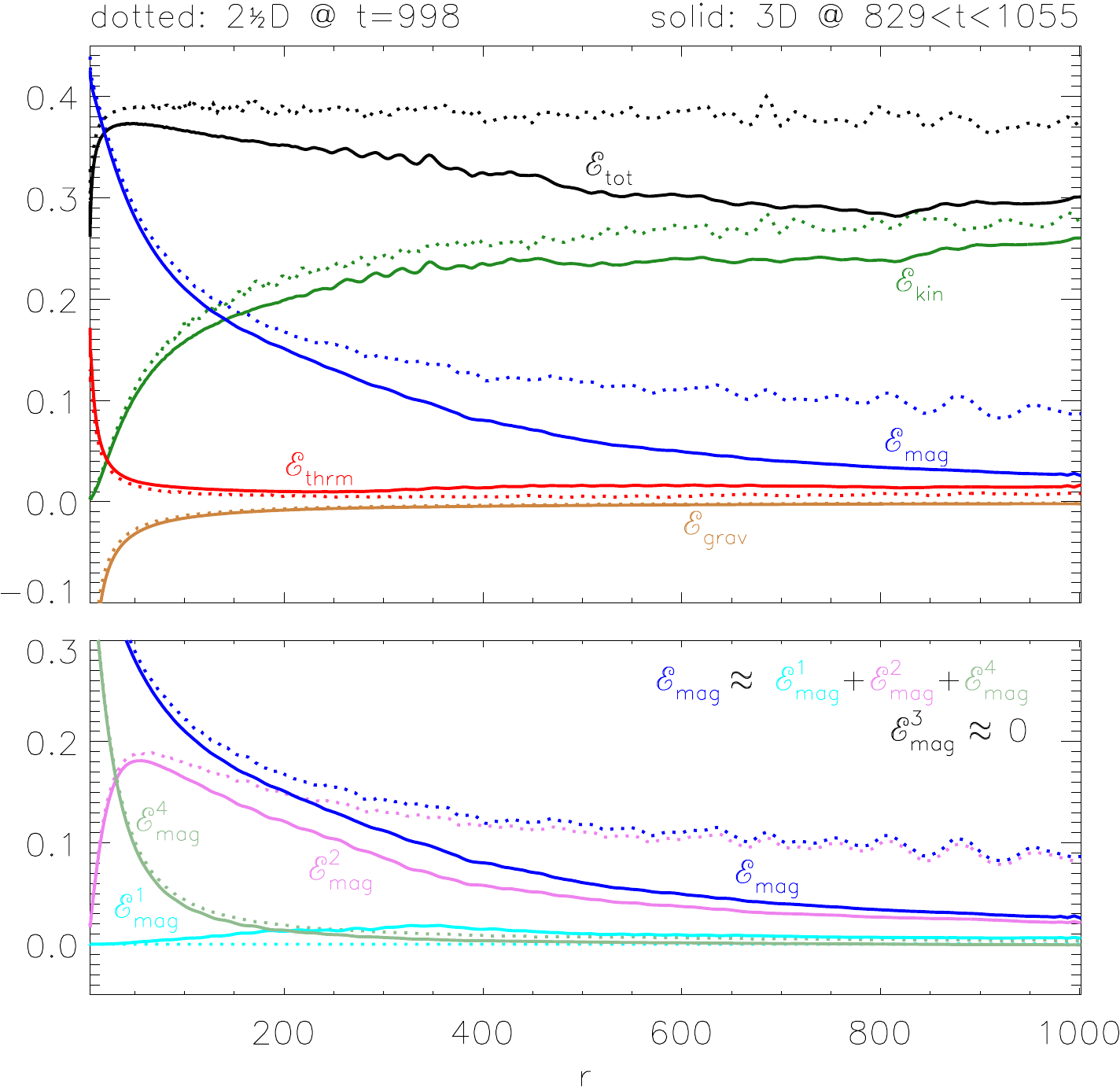}
\end{center}
\caption{Energy flow rates through the $r=\const$ surface. Magnetic enthalpy
(integral over Poynting flux, blue line) is reduced in the 3D simulation as
$B_\varphi^2$ decays as a consequence of instabilities.}
\label{fig:eflow}
\end{figure}

A comparison of the different kinds of energy flow rates gives information
about energy transformations taking place in the jet.   Integrating the radial
energy flux
\begin{equation}
    \underbrace{\frac{1}{2}\rho v^2 v_r}_\text{kinetic} + \underbrace{\frac{\gamma}{\gamma-1}p v_r}_\text{thermal enthalpy}
    + \underbrace{\rho \Phi v_r}_\text{grav. potential} + \underbrace{S_r}_\text{magnetic enthalpy}
\label{eq:eflux}
\end{equation}
over the $r=\const$ surface, we obtain the rate of energy
$\efrate_\text{tot}(r)$ that crosses the surface.  The individual components,
which we denote with $\efrate_\text{kin}$, $\efrate_\text{thrm}$,
$\efrate_\text{grav}$ and $\efrate_\text{mag}$, are plotted in
Fig.~\ref{fig:eflow}.  As the values are strongly fluctuating in the unstable
region, we averaged over time in the 3D case.  $\efrate_\text{mag}$ is reduced
by $80\%$ in the 2.5D simulation and by $94\%$ in the 3D simulation along the
simulated distance, the gap between the curves widens considerably where the 3D
jet exhibits instabilities.  We presume this to be caused by (numerical)
magnetic dissipation, which turns magnetic energy into heat, causing an
increase in $\efrate_\text{thrm}$. The environmental thermal energy is not
affected much.  As our calculations include an energy loss term (see
Sect.~\ref{sec:nummethods}), there is no one-to-one correspondence between the
dissipated energy and the increase in $\efrate_\text{thrm}$.

The final contribution of $\efrate_\text{mag}$ to $\efrate_\text{tot}$ is
$24\%$ in the 2.5D simulation and $9\%$ in the 3D simulation.
$\efrate_\text{kin}$ is up to about $10\%$ smaller in the 3D simulation and
does not show a dissipation-induced increase.  $\efrate_\text{kin}$ arises from
azimuthal and radial motion, with the relative share being highly similar in
the 2.5D and 3D case: the azimuthal contribution drops continuously to values
$\mathord{<}5\%$ at $r>200$ and the contribution from $v_\vartheta$ is
insignificant throughout.

$\efrate_\text{mag}$ can be decomposed further. The radial
component of the Poynting vector has 4 terms,
\begin{equation}
    S_r = \frac{1}{4\pi} \left( B_\vartheta^2 v_r + B_\varphi^2 v_r
    - B_\vartheta B_r v_\vartheta - B_\varphi B_r v_\varphi \right).
\label{eq:comps}
\end{equation}
Integrating these terms over the $r=\const$ surface gives the components [in
order of their appearance in Eq.~\eqref{eq:comps}] $\efrate_\text{mag}^1 \ldots
\efrate_\text{mag}^4$ which, added together, give $\efrate_\text{mag}$. These
components of the magnetic enthalpy flow rate are plotted in the lower panel of
Fig.~\ref{fig:eflow}.  At very small radii, the most important term is
$\efrate_\text{mag}^4$, the work done by the azimuthal flow against the
azimuthal component of magnetic stress.  Its contribution to
$\efrate_\text{mag}$ decreases rapidly, however, falling below $50\%$ at $r
\approx 30$ (i.e. well below the \Alfven surface, near the sonic one). It is
then taken over by $\efrate_\text{mag}^2$, which describes the flow of magnetic
enthalpy stored in the azimuthal field.  $\efrate_\text{mag}^1$ has minor
significance in the 3D simulation, which may be due to the perturbed toroidal
field having also a $B_\vartheta$ component (a rigid displacement of a pure
azimuthal field in the $r=\const$ plane introduces a non-azimuthal component).
$\efrate_\text{mag}^3$ is insignificant in both cases.  The strong decrease of
$\efrate_\text{mag}$ in the 3D simulation is caused by the decrease of
$\efrate_\text{mag}^2$.

We find a net outflow of magnetic enthalpy through the lateral boundaries of
the computational volume at the height of the jet front, with peak rates of
about $0.15$ in the 2.5D case and $0.05$ in the 3D case.  The outflow is
transient in the 2.5D case, vanishing quickly after the jet front leaves the
computational volume. In the 3D case, however, it turns into an inflow of the
order $-0.05$ which persists until the end of the simulation.  The energy in
the radial magnetic field increases correspondingly, mainly outside the jet at
$\vartheta>5.7\degree$.

\subsection{Magnetic field: poloidal vs. toroidal}
\label{sec:flux}

\begin{figure}[t]
\begin{center}
\includegraphics[width=\linewidth]{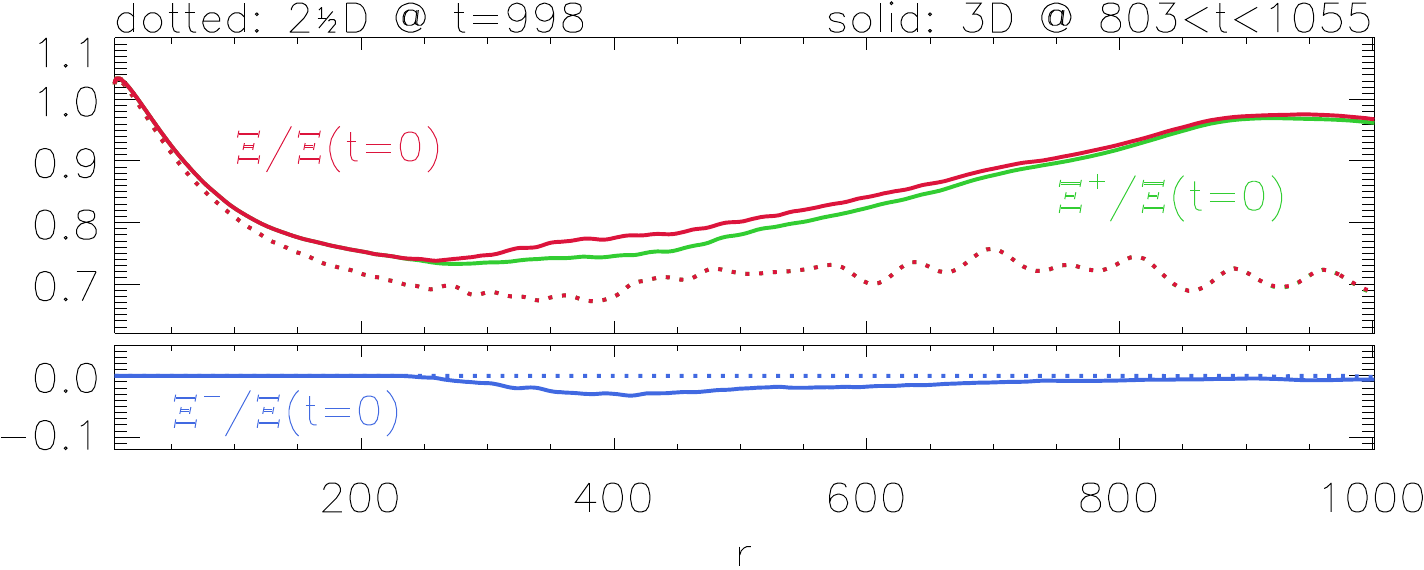}
\end{center}
\caption{Red: magnetic flux $\Xi(r)$  within $5.7\degree$ from the jet axis,
normalized by its initial value. Green: flux of $B_r^+ > 0$ ($B=B_r^+ +
B_r^-$), see Sect.~\ref{sec:flux}.}
\label{fig:magflux}
\end{figure}
\begin{figure}[t]
\begin{center}
\includegraphics[width=\linewidth]{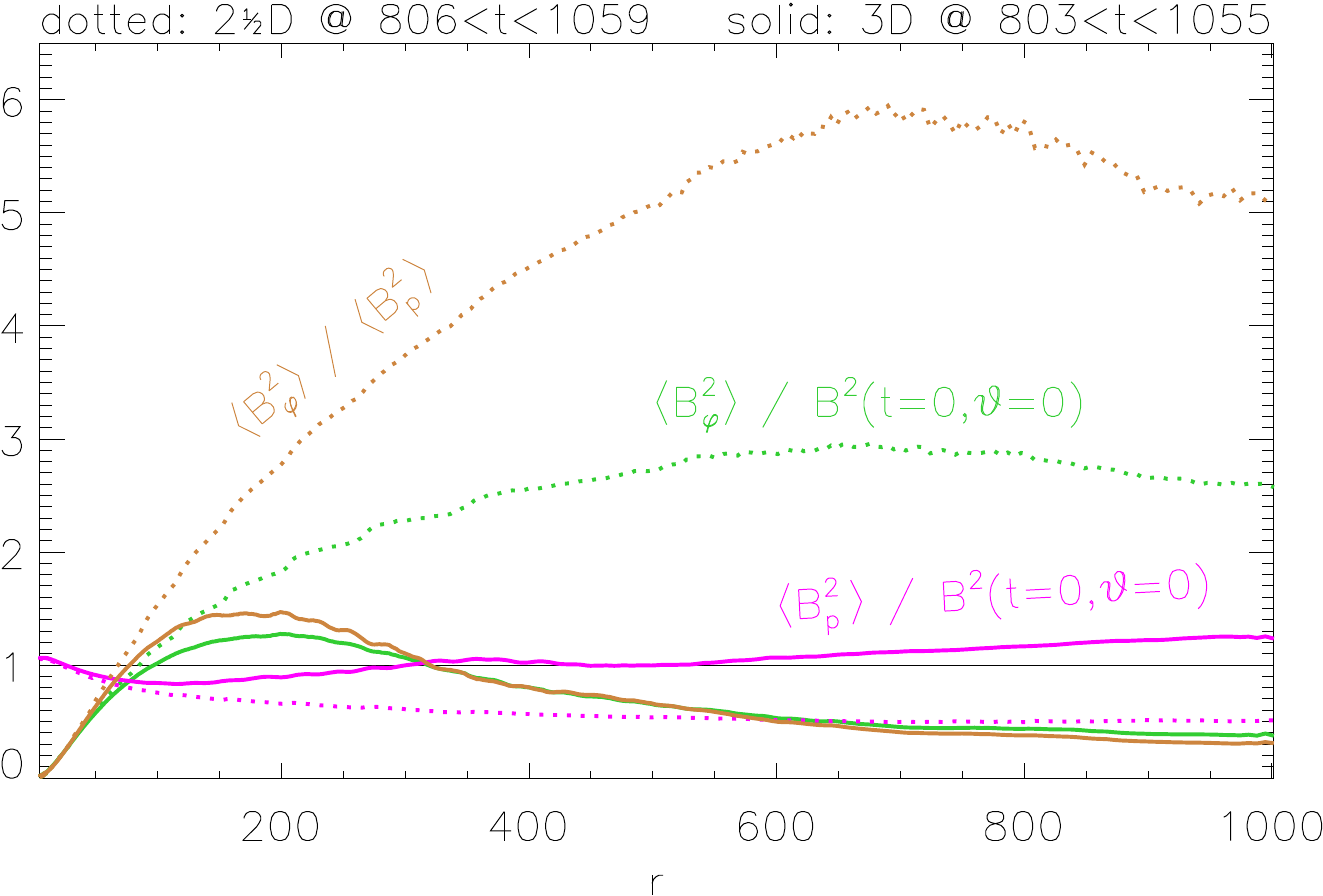}
\end{center}
\caption{Mean magnetic energies in the poloidal and toroidal fields within the
jet, normalized by the initial on-axis magnetic energy, and their ratio.
Without instability (2.5D), the field becomes predominantly toroidal at large
distances.  With instability (3D), it becomes predominantly poloidal.}
\label{fig:meanfields}
\end{figure}
\begin{figure}[t]
\begin{center}
\includegraphics[width=\linewidth]{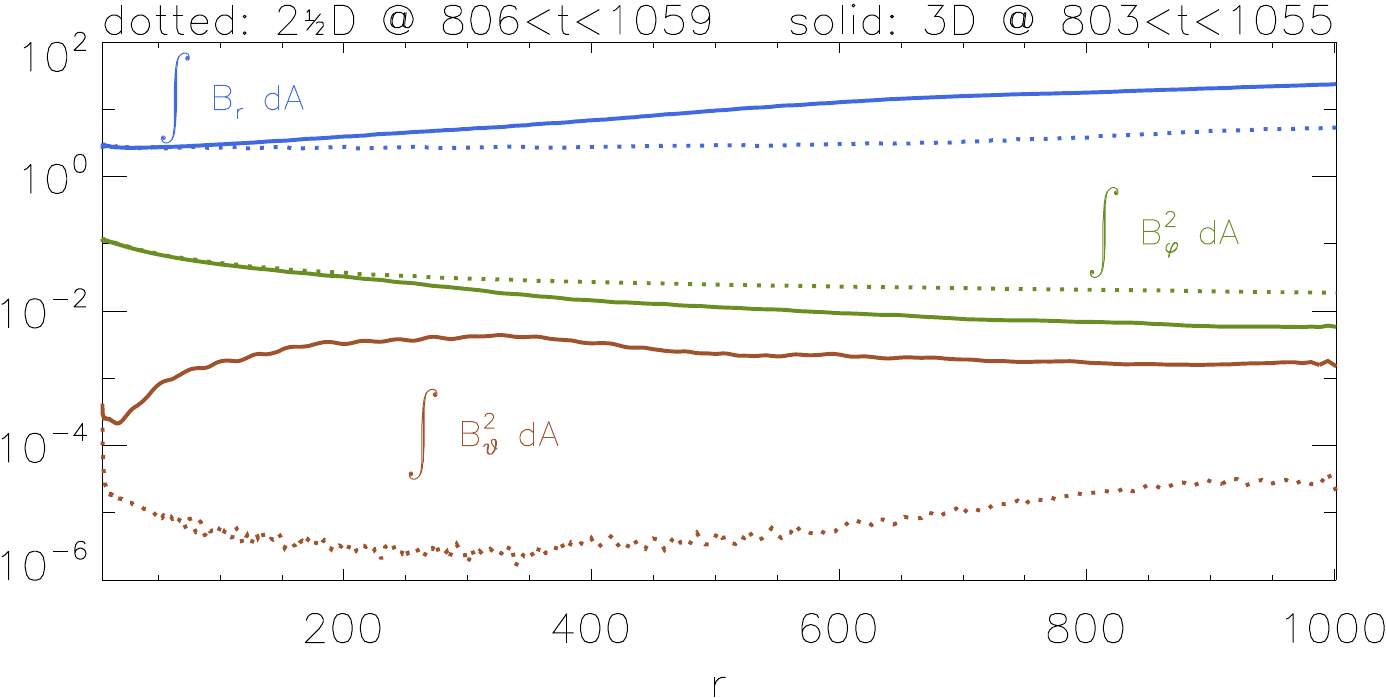}
\end{center}
\caption{Magnetic field components, integrated over the jet cross section, as a
function of distance, on a logarithmic scale.  In a spherically expanding
ballistic flow these quantities would be constant, see text.}
\label{fig:intfields}
\end{figure}
\begin{figure*}[t]
\sidecaption
\includegraphics[width=.29\linewidth,clip=true]{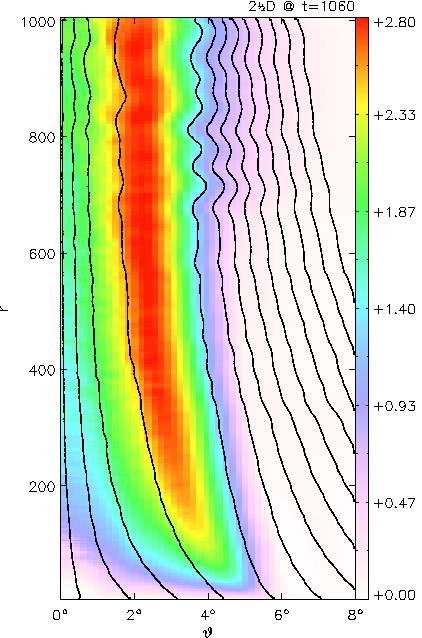}
\includegraphics[width=.29\linewidth,clip=true]{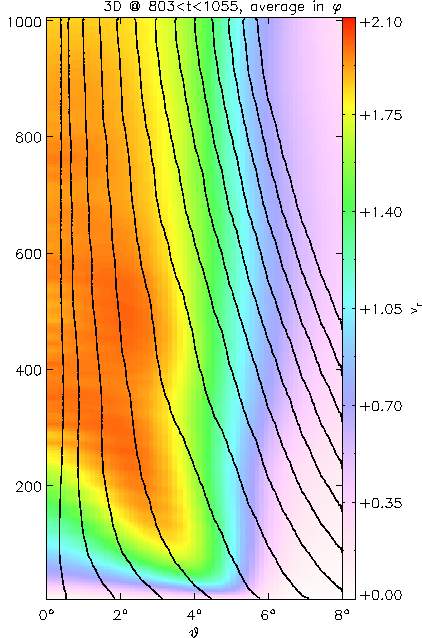}
\caption{Poloidal magnetic field lines and radial velocity field, plotted as a
function of direction $\vartheta$ (horizontal axis) and distance $r$ (vertical
axis); a conical magnetic surface would appear as a vertical line. The {\it
left-hand plot} shows a snapshot of the 2.5D case, the {\it right-hand plot}
shows the 3D case with the magnetic field and velocity being averaged in
azimuthal direction as well as time.  Both plots show the same field lines in
the sense that they start from the same direction at the lower boundary. The
angular separation between field lines in the outer, high-$\vartheta$ part of
the jet (here: between the 4th and 5th line from the left) increases noticeably
in the 2.5D case, leading to acceleration by the magnetic ``diverging nozzle''
effect there, see discussion.}
\label{fig:mflines}
\end{figure*}

The red line in Fig.~\ref{fig:magflux} shows the magnetic flux $\Xi$  contained 
within an angle  $\vartheta<5.7\degree$ from the axis,
\begin{equation}
    \Xi(r,t) \coloneqq \int_{\substack{r=\const\\\vartheta<5.7\degree}} B_r(t) \,\de A ,
\label{}
\end{equation}
divided by its initial value. For comparison, the green curve shows $\Xi^+$,
the flux of only those field lines that  have the same direction as the initial
field (the green and the red curves coincide in the 2.5D case).  Although the
2.5D jet fills only part of the $5.7\degree$ cone, it causes a significant
reduction of $\Xi$ . This reflects the lateral expansion of the jet due to the
pressure exerted by $B_\varphi$. There is less reduction in the 3D case, with
$\Xi$ being near the original value at very large distances.  The difference
becomes evident where the 3D jet is unstable, showing that it is due to the
dissipation of the toroidal field.  There is only a small difference between
the red and green curve in the 3D case, meaning that negative values of $B_r$
contribute little to the net flux. This shows that the toroidal field component
dissipates without producing much ``tangling'' of the field lines.

Fig.~\ref{fig:meanfields} compares the mean energy in the poloidal and toroidal
magnetic fields over the width of the jet as a function of distance, the width
of the jet being defined by a suitable velocity threshold.  The 3D and 2.5D
jets start out similar, with the mean poloidal and toroidal fields becoming
comparable near the \Alfven surface.  Beyond that distance, the mean toroidal
field energy increases more strongly in the 2.5D case, roughly proportional to
$r^{-2}$.  In the 3D case, the instability-induced destruction of the toroidal
field causes the slope to steepen substantially at $r \gtrsim 200$,
approximately as $\mean{B_\varphi^2} \sim r^{-3}$.  The mean poloidal field
energy, on the other hand, stays near the initial value in the 3D case while
decreasing somewhat in the 2.5D case.  Taking the ratio between the energies,
we find the magnetic field to be predominantly toroidal in the 2.5D case and
predominantly poloidal in the 3D case at large distances.

We also experimented with means of the form $\int X v_r \,\de A / \int v_r
\,\de A$, where the integral is performed over the whole $r=\const$ surface,
and obtained similar results.  Taking the mean within the static $\vartheta <
5.7\degree$ cone instead of using a velocity threshold yields a smaller value
for $\mean{B_\varphi^2}/\mean{B_\poloidal^2}$ in the 2.5D case, viz. $2\pm0.2$
for $r > 200$. This is because an angle of $5.7\degree$ includes more of the
environment of the jet (cf. Fig.~\ref{fig:border}).

In a ballistically expanding jet (constant velocity and opening angle) the
magnetic field components vary with distance $r$ as $B_r\sim r^{-2}$,
$B_\theta\sim B_\phi\sim r^{-1}$. Integrals over the width of the jet of $B_r$,
$B_\phi^2$ and $B_\theta^2$ are then constants.  These integrals are shown in
Fig.~\ref{fig:intfields}. The ballistic approximation works well in the 2.5D
case, for $B_r$ as well as for $B_\varphi$, if the acceleration region is
excluded. In the 3D case however, the integral of $B_r$ increases by about an
order of magnitude along the jet, while the integral of $B_\varphi^2$ decreases
over the entire range.

Fig.~\ref{fig:mflines} shows some poloidal magnetic field lines and the radial
velocity in the 2.5D and 3D jet. The latter was ``axisymmetrized'' by averaging
over the azimuthal coordinate $\varphi$. In the outer, high-$\vartheta$, part
of the 2.5D jet, where the toroidal field is especially strong and jet
acceleration most efficient, there is an increase of the angular separation
between the field lines which is absent in the 3D jet.  In other words, the
poloidal magnetic flux decreases locally faster with distance in the 2.5D case.

\subsection{Forces and powers}

\begin{figure}[t]
\begin{center}
\includegraphics[width=\linewidth]{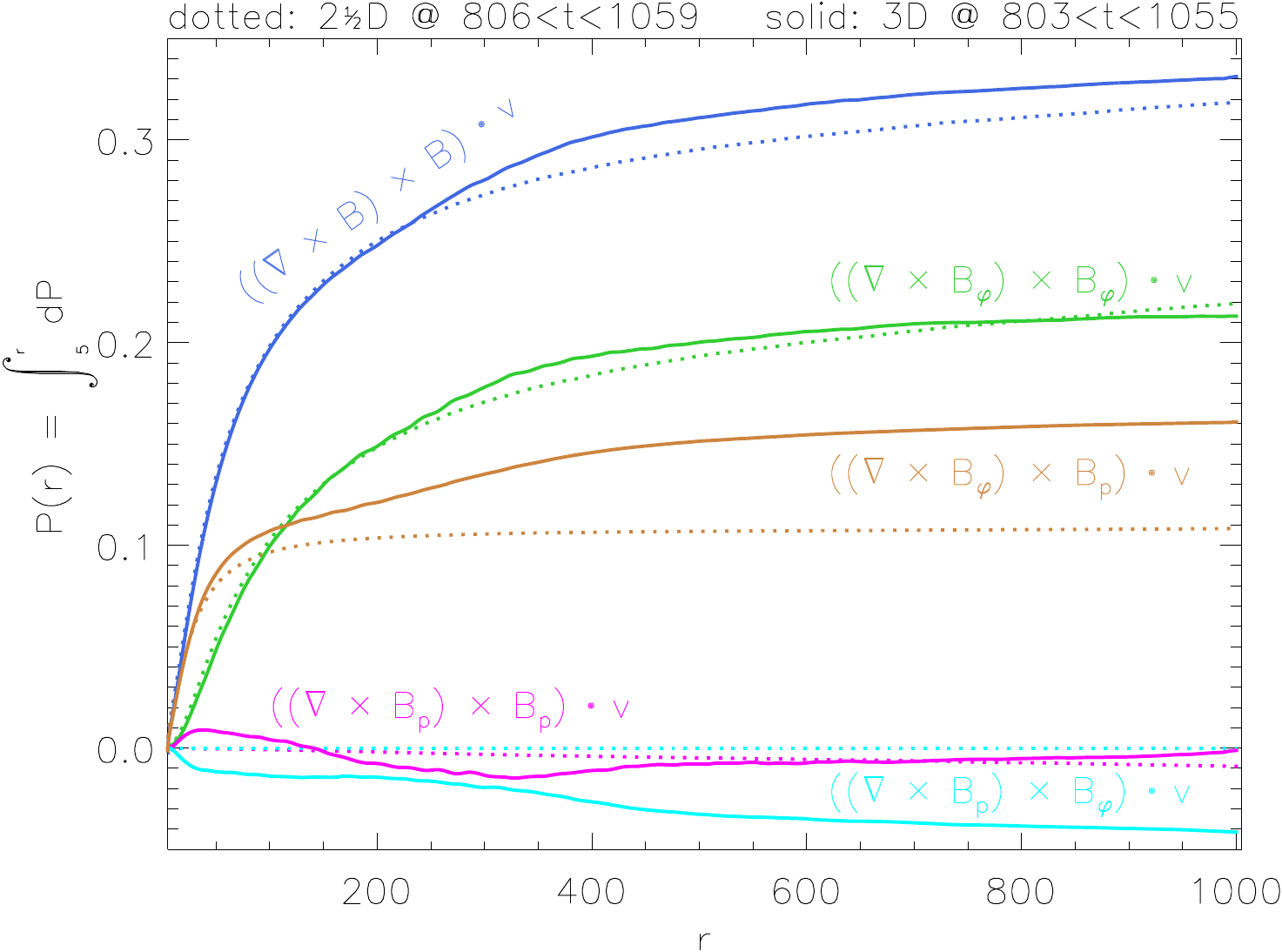}
\end{center}
\caption{Rate of work done in the volume delimited by the lower boundary and
$r$ by the components of the Lorentz force in the direction of the flow.  The
power delivered by the total Lorentz force is about the same in the 3D and 2.5D
cases.}
\label{fig:powers}
\end{figure}

To identify the accelerating forces that generate the kinetic energy flow
discussed in Sect.~\ref{sec:energy}, we compute $P = \int \vec{F} \cdot \vec{v}
\, \de V$, the instantaneous power (rate of work) delivered by a specific force
$\vec{F}$ in the direction of the flow in the integrated volume.  The
combination of gas pressure and gravitational forces, $-\Nabla p
-\rho\Nabla\Phi$, accounts for about one third of the power delivered by the
sum of all forces in the whole volume, the corresponding acceleration takes
place mainly below $r \approx 30$ (sonic surface).  The rest is accounted for
by the Lorentz force, which we decompose as follows:
\begin{align}
    \frac{1}{4\pi} (\Nabla \times \vec{B}) \times \vec{B} = \frac{1}{4\pi} \bigg[
        &(\Nabla \times \vec{B}_\poloidal) \times \vec{B}_\poloidal
        + (\Nabla \times \vec{B}_\varphi) \times \vec{B}_\varphi \nonumber\\
        + &(\Nabla \times \vec{B}_\poloidal) \times \vec{B}_\varphi
        + (\Nabla \times \vec{B}_\varphi) \times \vec{B}_\poloidal \bigg],
    \label{eq:lorentz}
\end{align}
with $\vec{B}_\varphi = B_\varphi \evarphi$ and $\vec{B}_\poloidal = \vec{B} -
\vec{B}_\varphi$.  The corresponding components of $P$ are plotted in
Fig.~\ref{fig:powers} as a function of the upper integral limit.

The last force in Eq.~\eqref{eq:lorentz} has only an azimuthal component. It is
important mainly below the \Alfven radius ($r \sim 60$), exerting a torque in
the same direction in which rotation is applied at the lower boundary.  The
next-to-last force vanishes in the axisymmetric case. In the general case, it
has only non-azimuthal components. Unlike the last force, it works against the
flow; the two forces largely cancel each other in the 3D simulation.  The
Lorentz force associated with $B_\varphi$ [second term in
Eq.~\eqref{eq:lorentz}] performs the same work in the 3D and 2.5D cases,
despite the steepening of the toroidal magnetic pressure profile caused by the
dissipation. The total power is also similar in the two cases, in agreement
with the similarity of the kinetic energy flows in Fig.~\ref{fig:eflow}.

The situation is a bit different if only the radial power $\int F_r v_r \, \de
V$ is taken into account. The Lorentz force associated with $B_\varphi$ does
approximately $20\%$ more work in the whole volume in the 3D case, the
additional power is delivered above $r \gtrsim 200$. The power of the net
force, however, is approximately the same in both cases.

\section{Summary and discussion}
\label{sec:discussion}

We have simulated jets generated by twisting a parabolically shaped large-scale
magnetic field in both 3D and axisymmetric 2.5D.  The shape of the initial
field reflects itself in a fair amount of jet collimation, with an opening
angle that decreases with distance, thus facilitating the growth of
instabilities.  The simulations cover the acceleration phase, where the jet
passes through the critical surfaces (sonic, \Alfven and fast magnetosonic) for
stationary MHD flows, as well as a substantial distance beyond these.  We thus
observe the onset of kink instabilities above the \Alfven surface in the 3D
simulation.  The instabilities disrupt the magnetic field structure.  They
cause magnetic dissipation, significantly reducing the toroidal field strength
and with it the flow rate of magnetic enthalpy (surface integral of the
Poynting flux) on a length scale of about 2--15 times the minimal \Alfven
distance in the jet.

A direct comparison of the 2.5D and 3D simulations reveals no significant
difference in the way the kinetic energy of the jet (integrated over its cross
section) increases with distance. However, the distribution of the kinetic
energy across the jet indicates differences in the acceleration process (cf.
Fig.~\ref{fig:mflines}).  In the axisymmetric flow, the acceleration is
restricted to magnetic surfaces that diverge from each other more rapidly than
in a flow with fixed opening angles.  This creates a ``magnetic nozzle'' effect
\citep{1994Begelman} restricted to a limited range of angles within the flow
(cf. discussion in \citealt{2008Spruit}).  A similar case of non-uniform
expansion has been found in relativistic flows by \citet{2008Tchekhovskoy}.  In
the 3D case, the (time averaged) acceleration is more uniform across the jet.
It thus seems that while the steepening of the magnetic pressure gradient
caused by the dissipation of the toroidal field leads to additional
acceleration in the 3D case, this is made up for by a more favorably
distributed poloidal magnetic flux in the 2.5D case which also enhances
acceleration.  In other words, there are different mechanism at work which
yield the same result. Whether it is coincidence that the two effects have
nearly the same strength is not clear.

The energy released by the dissipation of the toroidal magnetic field may be
radiated as light.  The yield depends on the details of the dissipation
(reconnection) and radiation mechanisms involved.  Since the magnetic energy
density ($B_\varphi^2/8\pi$) accounts for half of the Poynting flux
($B_\varphi^2 v_r/4\pi$), the present simulations suggest that the available
luminosity may be as much as $10\%$ of the initial magnetic enthalpy flow rate
(Fig.~\ref{fig:eflow}), provided that half of it is converted into kinetic
energy and the rest into light.  As dissipation is not a smooth process, the
emission will be stronger in some regions.  These would likely turn up as
bright knots in observations. Together with the wiggles caused by the kink
instabilities, the knots produce a structured appearance that is similar to
what has been observed in protostellar jets
\citep[e.g.][]{1996Heathcote,2001Reipurth}.  The knots in the mocked jet image
move with the flow, consistent with observational findings
\citep{1992Eisloeffel,2005Hartigan}. This provides an alternative to the
``internal shock'' interpretation usually invoked to explain jet knots
\citep[e.g.][]{1993Hartigan}.

The magnetic structure of a jet undergoes a dramatic change if it becomes
subject to violent kink instabilities. The formerly ordered helical structure
is largely destroyed and the poloidal field becomes the dominating field
component. However, magnetic flux conservation still implies that the poloidal
field declines faster with distance than the toroidal one.  The toroidal field
may thus, beyond the region covered by the here-presented simulations, become
again dominant. This could, in principle, lead to a resurgence of instabilities
until the \Alfven speed has dropped below the sideways expansion speed of the
jet (times a factor of order unity). From this point on the toroidal field is
effectively frozen in the flow (cf. discussion in Paper~1).

\subsection{Collimation and jet environment}

Collimation of jets is popularly attributed to the ``hoop stress'' in the
toroidal field component. This is misleading: though the stress contributed by
the toroidal field can compress the configuration near the axis (as observed in
simulations), it eventually has to be taken up by an external agent (for a more
extended discussion, see \citealt{2008Spruit}). In the simulations presented
above, as well as in other work, this agent is an external medium surrounding
the jet.  It is included mostly because of limitations of the codes used, since
the demands of conserving energy typically cause numerical instabilities at low
gas densities or low plasma-$\beta$.

The boundary between the jet and the external medium actually expands due to
pressure exerted by the toroidal field. The role of a (material) external
medium in confining the jet can be taken over by a magnetic field, if it is
able to transfer stress in the jet's toroidal field to the surface of the
accretion disk. One might imagine that a toroidal field extending around the
jet might serve this role. The high \Alfven speeds in this field, however,
would make it violently unstable to non-axisymmetric instabilities, as the
early history of magnetic configurations for controlled fusion (linear pinches)
testifies.  This obstacle does not become evident in the axisymmetric models in
the literature.  A more realistic possibility, proposed by \citet{1995Shu}, is
that the disk's poloidal field, assumed for launching the jet in the inner
regions, actually extends to much larger distances in the disk. Deformation of
the field can take up the lateral stress exerted by the jet. The collimation of
the jet would then be directly related to the properties of this poloidal
field.  High degrees of collimation would be most easily achieved in disks with
a large ratio of outer to inner disk radius \citep{1997Spruit}. Numerical
simulations at much lower plasma-$\beta$ than currently feasible would be
needed to study this form of ``poloidal collimation''.

As pointed out in the introduction, instabilities may take some distance to
travel to become effective. The results presented above show that a distance of
the order of $1000$ times the source size of the jet is needed to capture the
dissipation of the toroidal field. I surmise this to be the reason why the
effect of instabilities is not as noticeable in the works of
\citet{2006Anderson} and \citet{2003Ouyed}, rather than the lack of external
confinement of these winds.

\subsection{Disruption}

The possible presence of instabilities in jets sometimes raises concern about
disruption. A complete dissipation of the jet into its surroundings is possible
in principle through instabilities driven by the interaction of the jet with
its surroundings, for example by Kelvin-Helmholtz instabilities, or its
termination in a ``hot spot''. These processes extract their energy directly
from the bulk kinetic energy of the flow. The effect of internal instabilities
deriving from the free energy in the toroidal field is much less destructive,
since these mainly redistribute internal energy forms within the jet.  The
comparison between the 2.5D and the 3D cases shows that the 3D jet widens by
less than a factor 2 as a result of internal instabilities (cf.
Figs.~\ref{fig:border} and~\ref{fig:mflines}).  The non-axisymmetric nature of
the instabilities thus does cause some interaction with the environment, but
its consequences remain relatively benign.

\subsection{Cold flows}

Current 3D simulations are poorly equipped to handle flows in which the
magnetic or kinetic energy density, or both, dominate over the thermal energy
density.  In magnetically dominated flows driven from actual accretion disks,
the temperature of the plasma is often sufficiently low that magnetic energy
density dominates over plasma pressure already at a short distance from the
disk surface \citep{1982Blandford}.  As a consequence, the sonic point in the
present results is further away from the source than would be the case in e.g.
actual protostellar disks.  It is possible that the onset of the instabilities
in more strongly magnetically dominated flows would be faster, and their
consequences even stronger than in the simulations presented here. Codes
specially designed to handle such ``cold'' flows would be needed to verify
this.

\begin{acknowledgements}
The author thanks H.~C.~Spruit for fruitful discussions and a critical reading
of the manuscript, and M.~Obergaulinger for providing his MHD code.
\end{acknowledgements}

\bibliography{ref}

\end{document}